\documentstyle[12pt]{article}
\begin{document} 
\begin{flushright}
OITS 772\\
September 2005
\end{flushright}

\begin{flushleft}
\vskip-1.5cm
Talk given at Quark Matter 2005
\end{flushleft}
\vskip1cm

\begin{center}  {\Large {\bf Resolutions of Several Puzzles at
Intermediate $p_T$\\ and Recent Developments in Correlation}}
\vskip .75cm
 {\bf  Rudolph C. Hwa}
\vskip.5cm
{Institute of Theoretical Science and Department of
Physics\\ University of Oregon, Eugene, OR 97403-5203, USA}
\end{center}

\abstract{ 
Some of the puzzles on hadron production at intermediate $p_T$ found
at RHIC are explained as natural consequences of parton recombination. 
In that framework for hadronization the correlation among hadrons
produced in jets can be calculated.  Some new results on both near-side
and away-side jet structures are presented. }

\section{Introduction}
The traditional way of treating hadronization is by means of
fragmentation.  That has been found to lead to gross disagreement with
the data on hadron production in the $3 < p_T <8$
 GeV/c range at RHIC.  It has already been pointed out at Quark Matter 2004 that 
 the $p/\pi$ ratio can be understood in the
recombination model \cite{rf,rh}.  I now present several other puzzles
when interpreted in the fragmentation model, but they can readily be
resolved in the framework of parton recombination.  In that framework
correlations in jets can also be calculated. 
With or without triggers, correlation between two hadrons can be studied either in $p_T$ or in $\Delta\eta$ and $\Delta\phi$. The most recent result from work still in progress is the simulation of tracks of hard partons traversing the dense medium, showing the dip-bump structure on the away-side in $\Delta\phi$ but turning to a peak at higher parton momentum.

\section{Several puzzles and their resolutions}
The $p/\pi$ ratio in Au+Au collision was found to exceed $1$ at $p_T
\approx 3$ GeV/c \cite{sa}, a phenomenon that contradicts
fragmentation, since the fragmentation function for proton is much
less than that for pion.  Fragmentation is not an important process at low and intermediate $p_T$ because recombination that involves the soft thermal partons is the dominant process \cite{reco}.
When thermal-shower parton recombination is
considered, the data on the $p$ and $\pi$ spectra can be well
reproduced, and thus the $p/\pi$ ratio also \cite{hy}.

A similar phenomenon occurs in d+Au collisions.  The $A$ dependence
of the $p_T$ spectra is conventionally referred to as the Cronin effect,
\cite{jc} which has become synonymous to $p_T$ broadening by
multiple scattering in the intial state.  If the $A$ dependence is due to
the initial-state effect, and not to hadronization in the final state, then
that effect should be independent of the species of the hadron detected. 
But the data indicate otherwise \cite{fm,hy2}.  Again, in the
recombination model it has been possible to obtain
$R^p_{CP}>R^{\pi}_{CP}$ for d+Au collisions in agreement with the data
\cite{hy2}.

Another related phenomenon is the forward-backward asymmetry in d+Au collisions.  If the transverse broadening of the initial partons is
impotant, then there is more broadening in the forward ($F$) direction
than in the backward ($B$) direction.  Thus $B/F$ should be $< 1$.  But
the data show $B/F > 1$ for $1 < p_T < 5$ GeV/c \cite{ja}.  Since there are
more thermal partons in the $B$ direction than in the $F$ direction,
thermal-shower recombination leads naturally to $B/F > 1$.  In fact, the
data on $R_{CP}$ for $\eta = 0$ to $3.2$ \cite{ia} can be well
reproduced in the recombination model \cite{hyf}.  That is significant
because no new physics has been inserted in going from the backward to
the forward direction, in contrast to the approach based on saturation
physics \cite{jjm}.

The final puzzle we mention here is the difference of the associated
particle distributions (APD) in central $AA$ and $pp$ collisions on the
same side as the trigger \cite{ja2}.  Even with medium-modified 
fragmentation function  it
seems hard to accommodate a factor of 3 difference at $p_T \sim 1$
GeV/c.  That is readily obtained in the recombination model due to the
abundance of thermal partons at low $p_T$ in $AA$ collisions \cite{ht}.

\section{Correlations of hadrons in jets}

Since hadronization by recombination has been found to be so
successful in HIC, it is natural to exend the study to correlations among
hadrons in jets produced at RHIC.  There are various ways of studying
correlation; they can be divided into two classes.    One is to use a
trigger and study the APD in various variables.  The other is to treat two
hadrons on equal footing without designating one of them as trigger.  

\subsection{Correlations using trigger}

There exist data that show the $\Delta\eta$ and
$\Delta\phi$ dependences of the AP.  On the near side it is found that
a peak in $\Delta\eta$ sits on top of a flat background, called the
pedestal \cite{ja2}.  There is no such pedestal
in the $\Delta\phi$ distribution, mainly because of the subtraction
scheme that forces the APD to vanish at $|\Delta\phi| = 1$.  In \cite{ch}
the pedestal is interpreted as the result of the conversion from the hard
parton's energy loss to the thermal energy of the neighboring soft
partons, which in turn enhance the multiplicity of the AP.  Since the
energy loss of a hard parton is proportional to the distance it traverses
in the medium, lower trigger momentum allows the hard collision to
occur farther away from the surface than a trigger with higher
momentum.  Thus if our interpretation of the pedestal is correct, then
we expect the pedestal to disappear as the trigger momentum is
increased, since hard partons created nearer the surface lose less
energy on their way out.  There is some hint in the data that the pedestal
is diminished at higher trigger momentum \cite{dm}.

On the away side the $\Delta \phi$ distribution has stimulated
considerable interest because it gives information on the nature of jet
quenching.  The observation of a dip at $\Delta \phi = \pi$ and bumps
on the two sides at $\Delta \phi \sim \pi \pm 1$ when the trigger
momentum is less than 4 GeV/c \cite{ja2,wh} seems to support the idea
of collective response from the medium to the passage of a hard parton,
such as a shock wave.  However, such an explanation would encounter
difficulty, if the dip-bump structure disappears at higher trigger
momentum.  We have considered the problem by simulating parton
rescattering track-by-track based on a Gaussian distribiton of scattering
angle at discrete points in the dense medium, the distance between two
successive points being dependent on the average local density and the
parton momentum \cite{ch2}.  With a specific criterion for the
termination of a track, it is found that most trajectories directed toward
the center of the medium are totally absorbed.  The ones that emerge
have initial directions away from the center and undergo successive
scattering with deflections persistently farther away from the center;
they are the tracks that give rise to the bumps at $\Delta \phi$ away
from $\pi$, as observed in the data.  The lack of straight-through
trajectories causes the dip at $\Delta \phi = \pi$.  The absorbed tracks enhance the thermal partons and lead to higher multiplicity of the soft hadrons that lift the base of the dip. This scenario is
changed as the initial parton momentum is increased, since more
trajectories punch through the medium, resulting in a peak rather than
a dip.  There exists preliminary evidence for such a change in the
$\Delta \phi$ distribtion in the RHIC data as the trigger momentum is
increased \cite{dm2}. 

\subsection{Correlation without trigger}

The other class of correlation studies is the direct analysis of the correlation function $C_2(1,2)=\rho_2(1,2)-\rho_1(1)\rho_1(2)$, where $\rho_1$ and $\rho_2$ are the one- and two-particle distributions. If the focus is on the correlation between the $p_T$ values of the produced hadrons, there is no need for any subtraction of background beyond what is explicit in the definition of $C_2(1,2)$. On the other hand, if the interest is on the $\Delta\eta$ and $\Delta\phi$ distributions without treating one of the two hadrons as the trigger for reference, then autocorrelation is the quantity for investigation, as pioneered by Trainor and collaborators \cite{tpp,ja3}. We have some results on both of these types of correlations on the same side of a jet.

For correlation in $p_T$ our main result is that $C_2$ is negative for $p_T$ around 2 GeV/c in Au+Au collisions at 200 GeV \cite{ht2}. The reason for the anti-correlation is that the shower partons in a jet are anti-correlated, if it is assumed that the shower partons are dynamically independent, but kinematically constrained due to momentum conservation. Since thermal-shower recombination is the dominant component at intermediate $p_T$, the pion $C_2$ is therefore also negative. At $p_T$ higher than 3 GeV/c the $\rho_1(1)\rho_1(2)$ part of $C_2(1,2)$ becomes more severely damped due to the double appearance of the hard scattering cross section, one in each $\rho_1$, whereas $\rho_2$ has only one; hence, $C_2(1,2)$ becomes positive, though very small. No data are available yet to verify this prediction. If no dip is found in $C_2(1,2)$, it may mean that there exists dynamical correlation of the shower partons in a jet not yet incorporated.

Finally, we report here the recent result of our study of autocorrelation, $A(\eta_-,\phi_-)$, where $\eta_-$ and $\phi_-$ are the differences of $\eta$ and $\phi$ of the two particles, all other variables being integrated over. The difference angular variables are related to the angle $\chi$ between the momentum vectors of the two particles, which is the key link to the parton dynamics. Since the momentum of the shower parton and that of the pion that it forms with a thermal parton are collinear, the angle $\chi$ between the two pions is the same angle between the two corresponding shower partons, whose angular distributions relative to the jet axis can be described by a Gaussian with a width $\sigma$. Thus for every $\sigma$ it is possible to determine the autocorrelation distribution $A(\eta_-,\phi_-)$ \cite{ch3}. The data on $A(\eta_-,\phi_-)$, when they become available, can then be used to deduce phenomenologically the value of $\sigma$, thereby revealing a basic property of the jet cone.

\section{Conclusion}

	The recombination model has succeeded in resolving several puzzles at intermediate $p_T$ in HIC and makes possible detailed calculations of various expressions of correlations in jets. So far no features of the RHIC data have presented any difficulty for this mechanism of hadronization to explain. For very high $p_T$ in excess of 10 GeV/c shower-shower recombination becomes more important, and that is equivalent to fragmentation \cite{hy}. The region that best exhibits the interaction between a hard parton and the dense medium is at intermediate $p_T$.

\section*{Acknowledgment}
 I am grateful to C.\ B.\ Chiu, R.\ J.\ Fries, Z.\ Tan and C.\ B.\ Yang for collaborative work, the results of which are reported here.
 This work was supported, in part,  by the U.\ S.\ Department of Energy under
Grant No. DE-FG03-96ER40972.  
 
% \section*{References}

\end{document}